\documentclass[column,showpacs,showkeys,preprint,numbers,amsmath,amssymb]{revtex4}
\usepackage{graphicx}
\usepackage{amsmath}
\usepackage{subfigure}
\usepackage{ifpdf}
\usepackage{cases}

\begin{document}

\title{ \bf Pinned interface dipole-induced tunneling electroresistance in ferroelectric tunnel junctions }
\author{\bf  Yin-Zhong Wu}
\affiliation{Jiangsu Laboratory of Advanced Functional Materials and Physics Department, Changshu Institute of Technology, Changshu 215500,
China\footnote{Email address: yzwu@cslg.edu.cn}}

\begin{abstract}
Based on the structure predicted in a ferroelectric tunnel junction(FTJ) in the resent density functional theory study[Phys.~Rev.~B $\bf{85}$
125407(2012)], we investigate the electron transport through the FTJ with asymmetric interfaces, i.e., one interface dipole is pinned and the
other interface dipole is switchable. Tuneling electroresistance(TER) can be induced due to the nonswitchable interface dipole in FTJs with
symmetric electrodes. Compared with the dependence relationship between TER and the polarization of switchable interface, TER is not sensitive
to the variation of the polarization of pinned interface. A large TER can be achieved when the pinned polarization points to the ferroelectric
film and low interface dielectric constants. In addition, effect of electrode on TER in the structure is also discussed.

\end{abstract}

\pacs{73.40.Gk, 77.55.fe, 77.80.bn} \keywords{Pinned interfacial dipole; tunneling electroresistance; ferroelectric tunnel junction;}

\maketitle
\section{Introduction}
The existence of ferroelectricity in nanometer-thick films makes it possible to use ferroelectrics as barrier in tunnel
junctions\cite{Science2006}, FTJ with polarization switching show important electronic transport properties, such as giant
TER\cite{Nature2009,APL2010}, solid-state memories\cite{natnanot2011}. Furthermore, switchable polarization of the tunnel barrier allows for
controlling of the transport spin polarization if the electrodes are ferromagnetic\cite{science2010,natmater2012}.

In theory, Tsymbal's group proposed that the surface charges in the ferroelectric are not completely screened by the adjacent metals and
therefore the depolarizing electric field in the ferroelectric is not zero. If a FTJ is made of asymmetric metal electrodes which have different
screening lengths, this lead to the asymmetry in the potential profile for the opposite polarization directions. Thus, the potential seen by
transport electrons changes with the polarization reversal which lead to a giant TER effects\cite{Science2006,PRL2005}. In general, the
interface is intrinsic and inevitable in the metal-ferroelectric-metal structure during preparing process. The presence of interfaces will
reduce the symmetry of a FTJ, and may produce electric dipoles at the interface. Duan\cite{NanoL2006Duan}studied ultrathin KNbO$_{3}$
ferroelectric films placed between metal electrodes, either SrRuO$_{3}$(SRO) or Pt, it is found that the bounding at the ferroelectric-metal
interfaces imposes severe constrains on the displacement of atoms, and induces an fixed interface dipole moment, which is electrode dependent
and non-switchable. Stengel and Spaldin\cite{Nature2006} demostrated the existence of a dielectric dead layer in SRO/SrTiO$_{3}$/SRO nanoscale
capacitor, which is only one or two atomic monolayers at the interface. Based on first-principles and model calculations, Wang\cite{PEB82}
investigated a ferroelectric dead layer near the interfaces that is nonswitchable, and predicted mechanism for the dead layer at the interface
controls the critical thickness for ferroelectricity in systems with polar interfaces. Liu\cite{PRBLiu}found that a BaO/RuO$_{2}$ interface
termination sequence in SRO/BaTiO$_3$/SRO epitaxial heterostructure grown on SrTiO$_{3}$(STO) can lead to a nonswitchable polarization state for
thin BaTiO$_{3}$(BTO) films due to a fixed interface dipole, while the interfacial polarization at the TiO$_{2}$/SrO interface is switchable.
And such asymmetric interfaces may destroy the stability of one of the polarization states, making the system only monostable in zero applied
field and therefore nonferroelectrics. Their first-principles and phenomenological modeling prediction and experimental measurments\cite{AM2012}
confirm that switchable ferroelectric polarization can be achieved by inserting a thin STO layer at BaO/RuO$_{2}$ interface to eliminate the
nonswitchable interface dipole. Previous studies are focused on the interfacial effect on the polarization\cite{PEB82} and critical
thickness\cite{PRBLiu} of the ferroelectric barrier. Recently, we have investigated electron transport in the FTJ with nonswitchable interface
polarizations at both the left and right metal-ferroelectrics interfaces\cite{myAPL2011}. However, the transport properties of FTJ with
asymmetric interface, i.e., one interface dipole is switchable and another interface dipole is nonswitchable, has not been studied, and such a
structure has been predicted by first-principles calculations in epitaxial SRO/BTO/SRO heterostructrure\cite{PRBLiu}. In this paper, effects of
electrode, interface polarization and interface dielectric constant on the TER in FTJs with a pinned interface dipole are investigated in
detail. We obtain that the pinned interface dipole can induce electroresistance in a FTJ with symmetric electrodes. Furthermore, comparison
between switchable and nonswitchable interface dipoles on TER are also given.

\begin{figure}
\vskip -0.9cm
\includegraphics[width=0.45\textwidth]{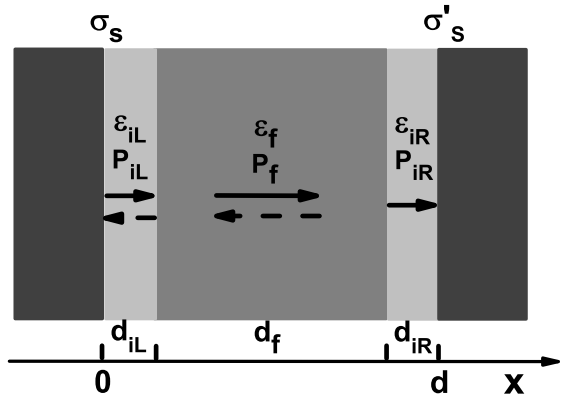}
\vskip -0.5cm
  \caption{\it{Sketch of the FTJ with a pinned interfacial dipole P$_{iR}$, whose direction can be arbitrarily preassigned. P$_{iL}$ and P$_{f}$ stand for the polarization of the left interface and the medial ferroelectric film,
  and they can be reversed by an external electric field.}}
\vskip 0.6cm
\end{figure}

\section{Model and Theory}

The structure of a FTJ with two interfaces between the metal electrodes and ferroelectric barrier is shown in Fig.~1. It is assumed that the
polarizations of the interface and ferroelectric barrier are perpendicular to the electrode, which can be realized by designing the misfit
compression stress from the substrate. Similar to the results of first-principles calculations in SRO/BTO/SRO nanostructure, we assume the
dipole in the right interface is pinned, and its direction  can be arbitrarily selected as pointing to the left or right. Once selected, its
direction cannot be switched under applied electric fields. The dashed arrows in Fig.~1 indicate that the corresponding polarizations can be
reversed by an applied field. We do not consider inhomogeneous polarizations throughout the barrier for simplicity, and the interface layer is
selected as one unit cell(d$_{iL}$=d$_{iR}$=4\AA)\cite{PRB4,myAPL2011}.

According to the Thomas-Fermi model, the screening potential within
the left and right electrodes can be given by
\begin{equation}
  \varphi(x)=
  \Bigg\{\begin{array}{lc}
   -\frac{\sigma_{s}\lambda_{L}}{\varepsilon_{L}}e^{x/\lambda_{L}}, \hskip 1 cm & x\leq 0,\\
   \frac{\sigma_{s}\lambda_{R}}{\epsilon_{R}}e^{-(x-d)/\lambda_{R}}, \hskip 1 cm & x\geq d.\\
  \end{array}
\end{equation}

Here $\lambda_{L}$($\lambda_{R}$) is the Thomas-Fermi screening length of the left(right) electrode, $\epsilon_{L}$ and $\epsilon_{R}$ are the
dielectric constants of the left and right electrodes. $\sigma_{S}$ stands for the screening charge per unit area in the metal electrode which
is the same in the left and right electrodes due to the charge conservation condition. The continuity conditions of electric displacement and
electric potential at the boundaries are
\begin{subnumcases}{}
  \begin{array}{ll}
 \epsilon_{iL}E_{iL}+P_{iL}=\sigma_{s},\hskip 1 cm & x=0,\\
 \epsilon_{f}E_{f}+P_{f}-(\epsilon_{iL}E_{iL}+P_{iL})=0,\hskip 1 cm &x=d_{iL},\\
 \epsilon_{iR}E_{iR}+P_{iR}-(\epsilon_{f}E_{f}+P_{f})=0,\hskip 1 cm &x=d_{iL}+d_f,\\
 \epsilon_{iR}E_{iR}+P_{iR}=\sigma_{s},\hskip 1 cm &x=d_{iL}+d_f+d_{iR},\\
 \end{array}\\
E_{iL}d_{iL}+E_fd_f+E_{iR}d_{iR}=\varphi(0)-\varphi(d),
\end{subnumcases}
where $E_{iL}$, $E_{f}$ and $E_{iR}$ are the depolarization fields in the left interface, the barrier and the right interface, respectively.
$P_{iL}$, $P_{iR}$ and $P_{f}$ are the polarization of the interfaces and barrier. From Eqs.~(1) and (2), the depolarization fields and then the
electrostatic profile $\varphi$(x) ($0<x<d$) are obtained as follows

\begin{equation}
 \begin{array}{ccc}
 E_{iL}^{d}&=&\frac{\sigma_{s}-P_{iL}} {\epsilon_{iL}},\\
 E_{f}^{d}&=&\frac{\sigma_{s}-P_{f}}{\epsilon_{f}},\\
 E_{iR}^{d}&=&\frac{\sigma_{s}-P_{iR}}{\epsilon_{iR}},\\
 \end{array}
\end{equation}
and
\begin{equation}
   \varphi(x)=\Bigg\{
   \begin{array}{ll}
   \varphi(0)$-$E_{iL}^{d}x,&0<x\leq d_{iL},\\
   \varphi(d_{iL})$-$E_{f}^{d}(x$-$d_{iL}),&d_{iL}<x\leq d_{iL}$+$d_{f},\\
   \varphi(d_{iL}$+$d_f)$-$E_{iR}^{d}(x$-$d_{iL}$-$d_{f})$,$&d_{iL}$+$d_{f}<x\leq d,
   \end{array}
\end{equation}
where
\begin{equation}
\sigma_{s}= \frac{\frac{P_{iL}d_{iL}}{\epsilon_{iL}}+\frac{P_{f}d_{f}}{\epsilon_{f}}+\frac{P_{iR}d_{iR}}{\epsilon_{iR}}}
{\frac{\lambda_{L}}{\epsilon_{L}}+\frac{d_{iL}}{\epsilon_{iL}}+\frac{d_{f}}{\epsilon_{f}}+\frac{d_{iR}}{\epsilon_{iR}}
+\frac{\lambda_{R}}{\epsilon_{R}}}. \end{equation}

 The overall potential profile U(x) seen by transport electrons across the junction
is the superposition of the electrostatic potential -e$\varphi$(x), the electronic potential which determines the bottom of the bands in the two
electrodes with respect to the Fermi energy $E_{F}$, and the potential barrier created by the interfaces and barrier. It is assumed that the
interfaces potential and the barrier potential have a rectangular shape of height $U_{iL}$, $U_{iR}$ and $U_{f}$ with respect to the
$E_{F}$\cite{PRL2005}.

At a small bias voltage the conductance per unit area of the junction can be obtained using Landauer formula\cite{book1,book2}
\begin{equation}
G=\frac{2e^{2}}{h}\int{\frac{dk_{\parallel}}{(2\pi)^{2}}T(E_{F},k_{\parallel})},
\end{equation}
where $T(E_{F},k_{||})$ is the transmission coefficient at the Fermi energy for a given transverse wave vector $k_{||}$, and can be calculated
by solving the schr\"{o}dinger equation for an electron tunneling through the potential barrier $U(x)$ within the formalism of transfer matrix
\cite{APL3,Spacecharge}. It is necessary to note again that, as show in Fig.~1, the dipole $P_{iL}$ within the left interface and the
spontaneous polarization $P_{f}$ in the ferroelectric barrier are always in the same direction, and their direction can be reversed under the
action of an external electric field, while the dipole $P_{iR}$ in the right interface is nonswitchable. In this paper, the right direction is
defined as the positive polarization direction, and the TER ratio is defined as TER=$\frac{G_R-G_L}{G_L}$, where $G_L$ and $G_R$ are
conductances of a FTJ for polarization in the barrier pointing left and right, respectively. It is well known that the larger the value TER is,
the better the performance of FTJs.

\section{ Results and Discussions}

Firstly, the TER of a FTJ with a pinned interface dipole and symmetric electrodes is numerically calculated. Pt and SRO electrodes are used in
our study to simulate effect of electrodes on TER, as well as the effect of nonswitchable interface dipole on TER. The screening length of
Pt(SRO) is selected as 0.4\AA(0.75\AA), and the dielectric constants of Pt and SRO electrodes takes the values $\epsilon_{0}$ and
$8.85\epsilon_{0}$, respectively\cite{Spacecharge}. The Fermi energy is chosen as $E_{F}=3.0eV$, and the barrier heights for the interface and
FE barrier are taken as $U_{iL}=U_{iR}=0.6eV$ and $U_{f}=0.6eV$\cite{myAPL2011}. The polarization, the dielectric constant and the thickness  of
the ferroelectric film are chosen as $P_{f}=0.26C/m^2$, $\epsilon_f=90\epsilon_0$ and $d_f=2nm$ in this paper.
\begin{figure}
\centering \subfigure[]{\includegraphics[width=0.45\textwidth]{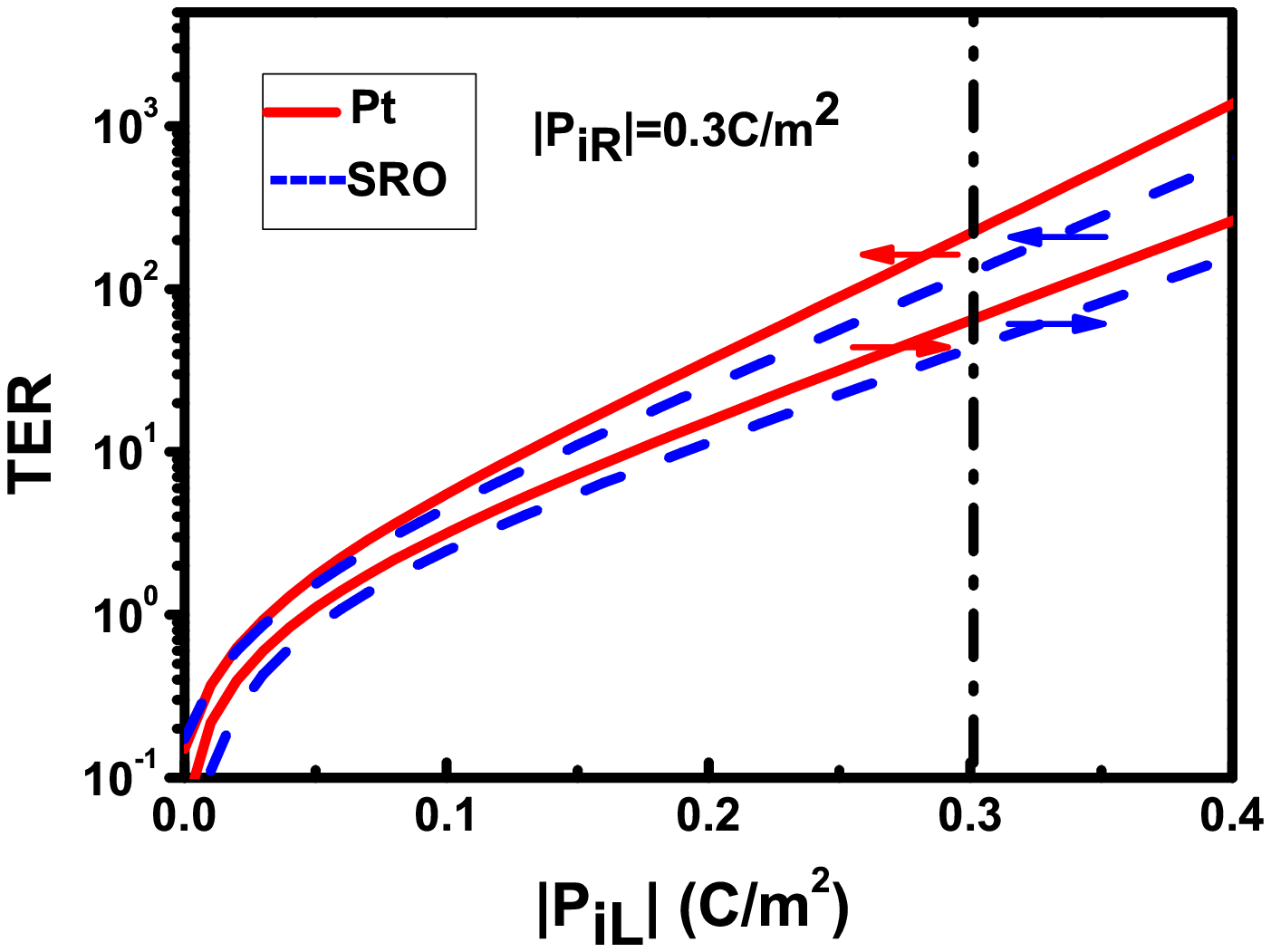}}
\subfigure[]{\includegraphics[width=0.45\textwidth]{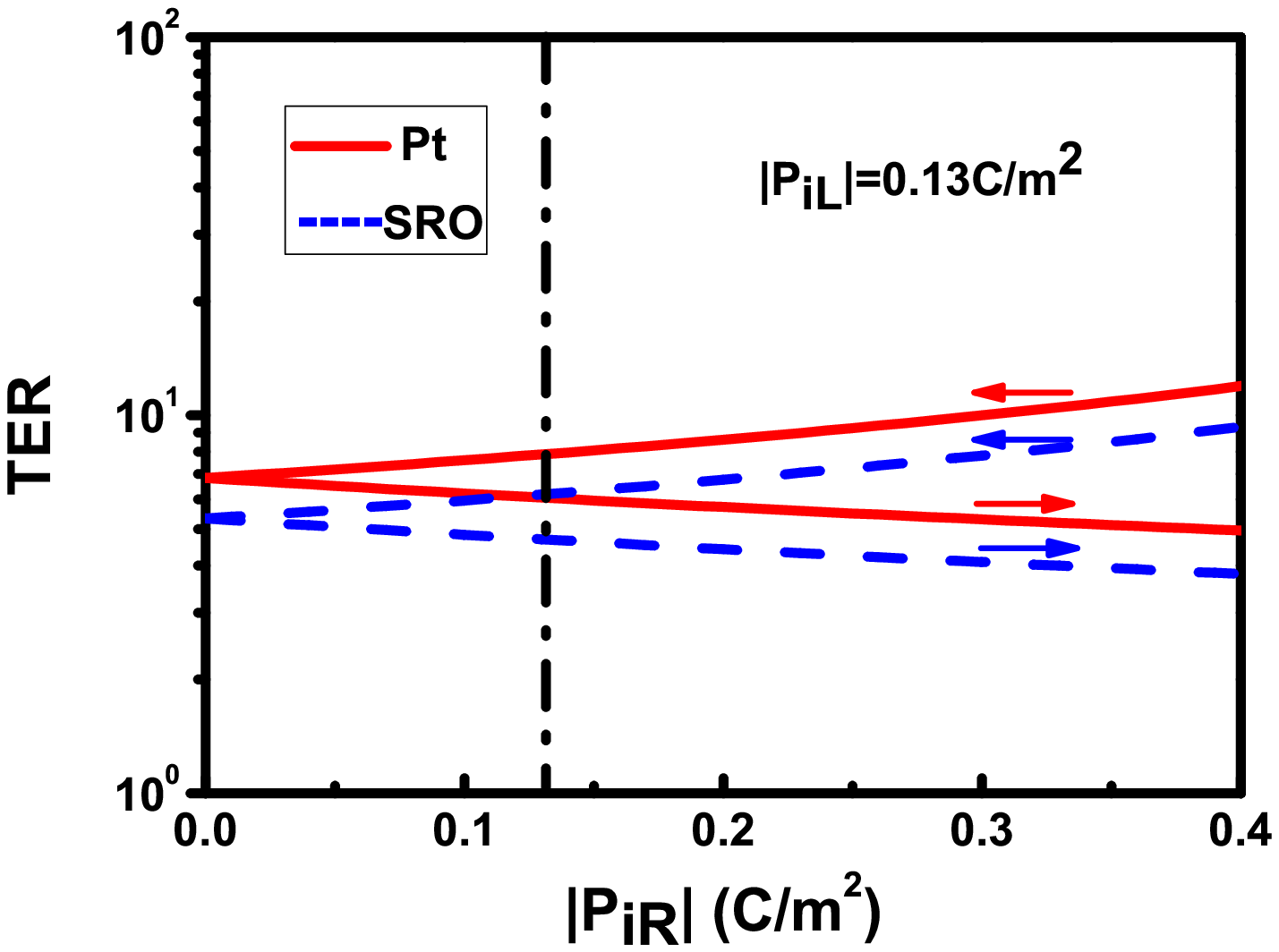}}
\caption{ \it{(Color online) TER as a function of the magnitude of (a) $P_{iL}$ and (b) $P_{iR}$. Solid lines denote the TER for Pt electrodes,
dashed lines for SRO electrodes, and dash-dot line is drawn to indicate where the magnitude of the two interfaces are equal. Arrows represent
the direction of nonswitchable $P_{iR}$, which is assumed to preassign only one of the two directions.}} \label{Fig.2}
\end{figure}

In Fig.~2, TER as a function of the polarizations of switchable interface dipole $P_{iL}$ and pinned interface dipole $P_{iR}$ are plotted. The
arrows near the lines stand for the direction of $P_{iR}$, which is preassigned. The vertical dash-dot line in Fig.~2 denotes where the
magnitude of polarization of the left interface and the right interface are equal, however, the left interface dipole is switchable, and the
right one is nonswitchable. If the two interfaces have the same polarization and both of them are switchable, then TER of the corresponding FTJ
will be zero\cite{PRL2005}. From the intersection of dash-dot line and TER curves in Fig.~2, one can see that TER is not zero. Therefore, TER
can be induced by asymmetric interface dipoles, such as the asymmetry in switching. Moreover, TER always increases with the increase of the
magnitude of switchable dipole $P_{iL}$, while TER increases with increase of $P_{iR}$ only when the nonswitchbale dipole $P_{iR}$ pointing to
the barrier(left), and TER decreases as $P_{iR}$ increases when $P_{iR}$ points to the electrode(right). Compared Fig.~2(a) with 2(b), we find
that TER is more sensitive to the change of $P_{iL}$ than the change of $P_{iR}$, it is because that the left interface dipole $P_{iL}$ is
switchable and has the same direction with $P_{f}$, therefore the increase of $P_{iL}$ is equal to enhance the effective spontaneous
polarization of the FTJ, then, results in an increase of TER. Under the same conditions, the nonswitchable dipole $P_{iR}$ pointing to barrier
favors a large TER, which is consistent with our previous results\cite{myAPL2011}. As the effect of electrode is concerned, the FTJ with Pt
electrodes will have a large TER than that with SRO electrodes when the interfacial dielectric constants take the average value between the
dielectric constant of electrode and that of ferroelectric film. Here, $\epsilon_{iL}=\epsilon_{iR}=45.5\epsilon_0$ for Pt electrodes and
$\epsilon_{iL}=\epsilon_{iR}=49\epsilon_0$ for SRO electrodes in Fig.~2.
\begin{figure}
\centering
\subfigure[]{\includegraphics[width=0.45\textwidth]{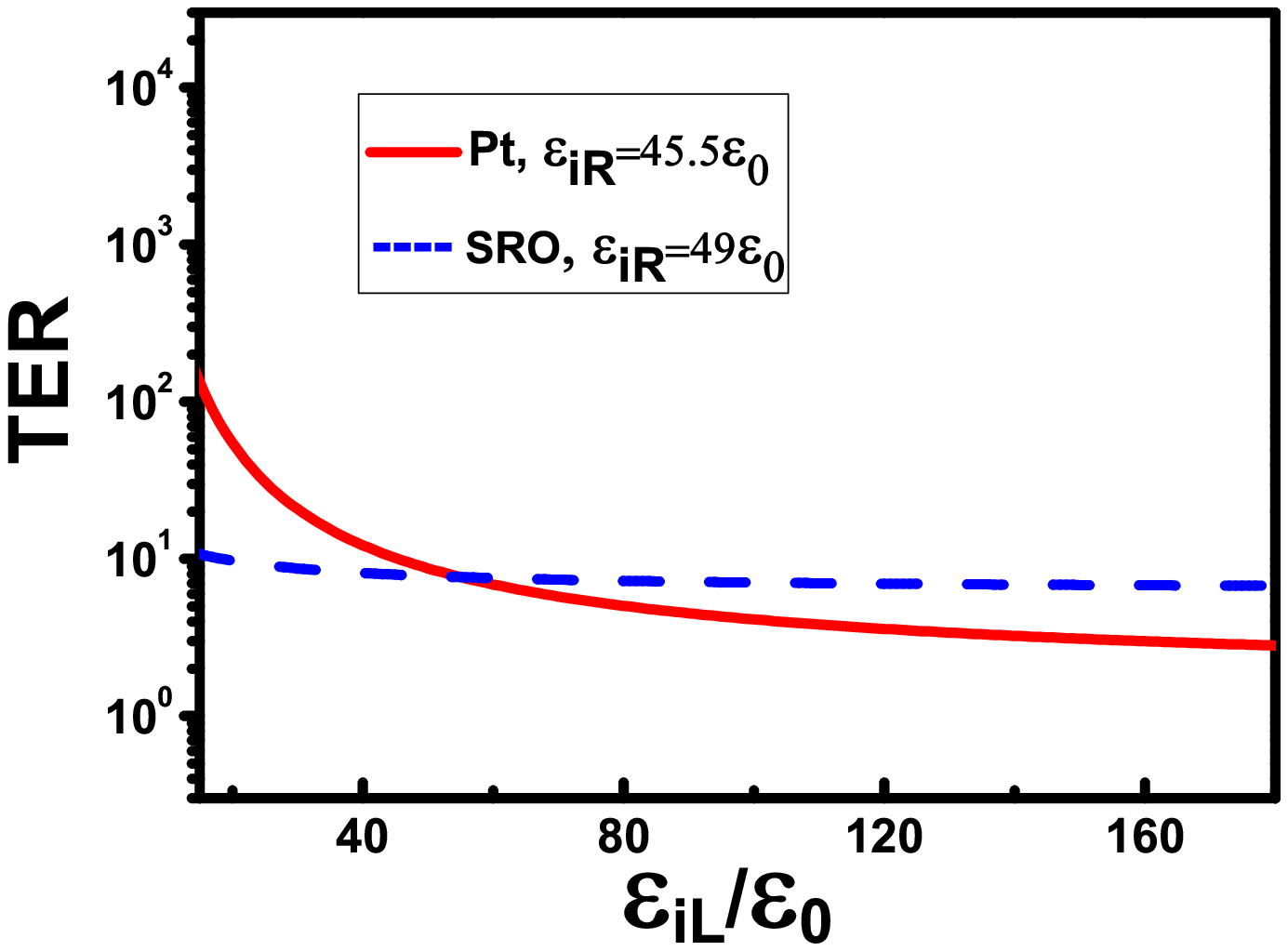}}
 \mbox{\hspace{0.5cm}}
\subfigure[]{\includegraphics[width=0.45\textwidth]{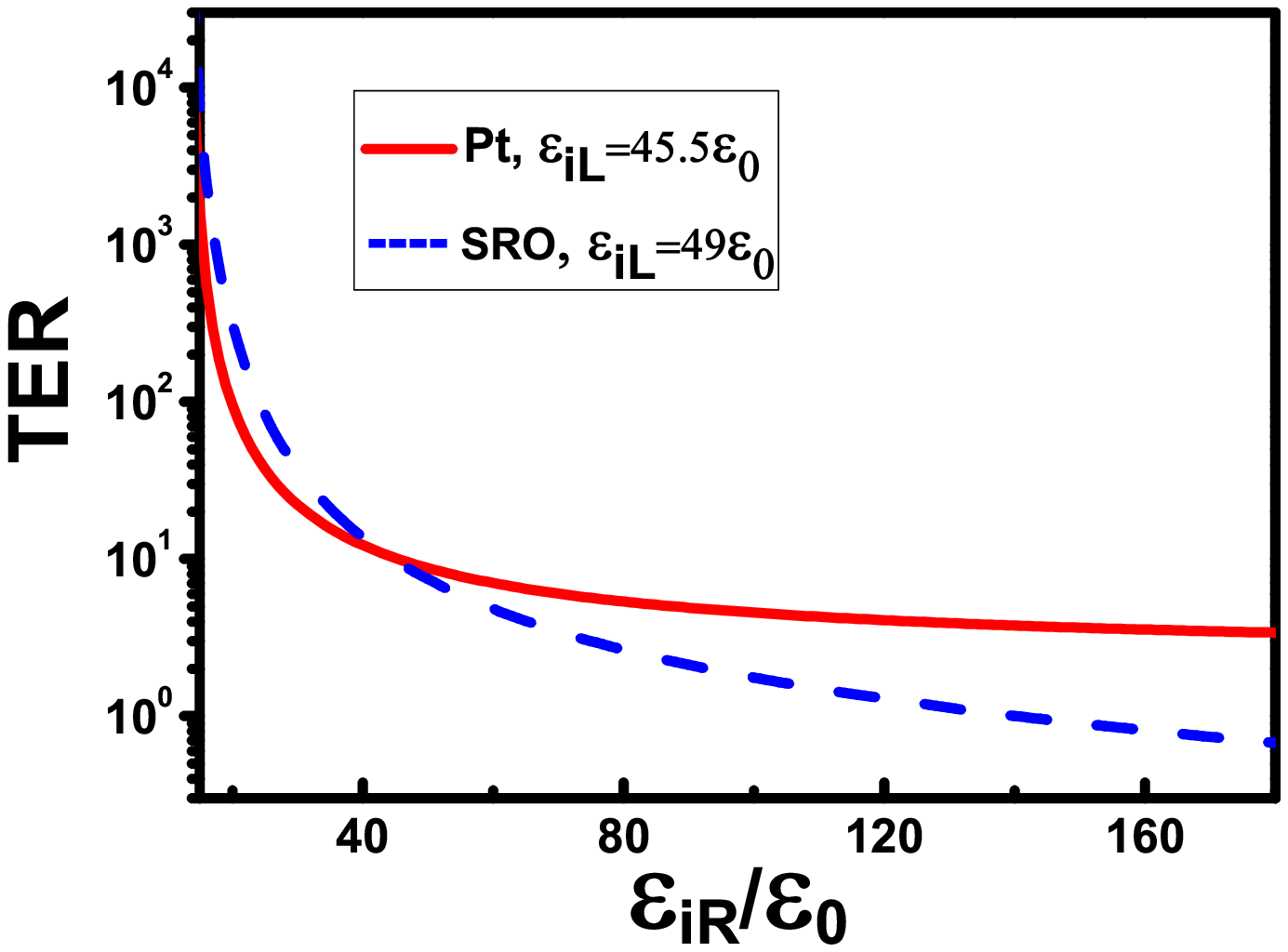}}
\\
\subfigure[]{\includegraphics[width=0.45\textwidth]{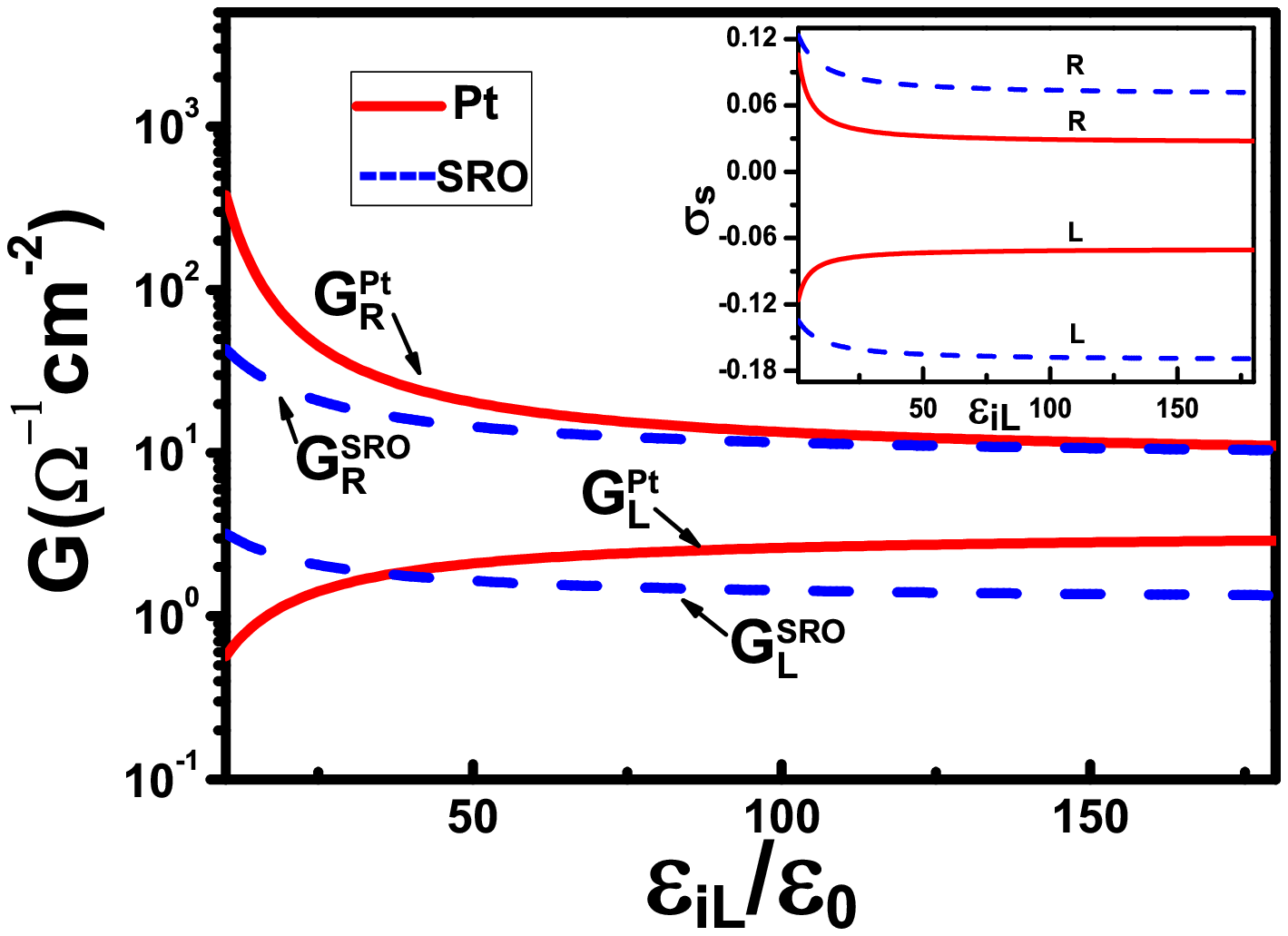}}
\mbox{\hspace{0.5cm}}
\subfigure[]{\includegraphics[width=0.45\textwidth]{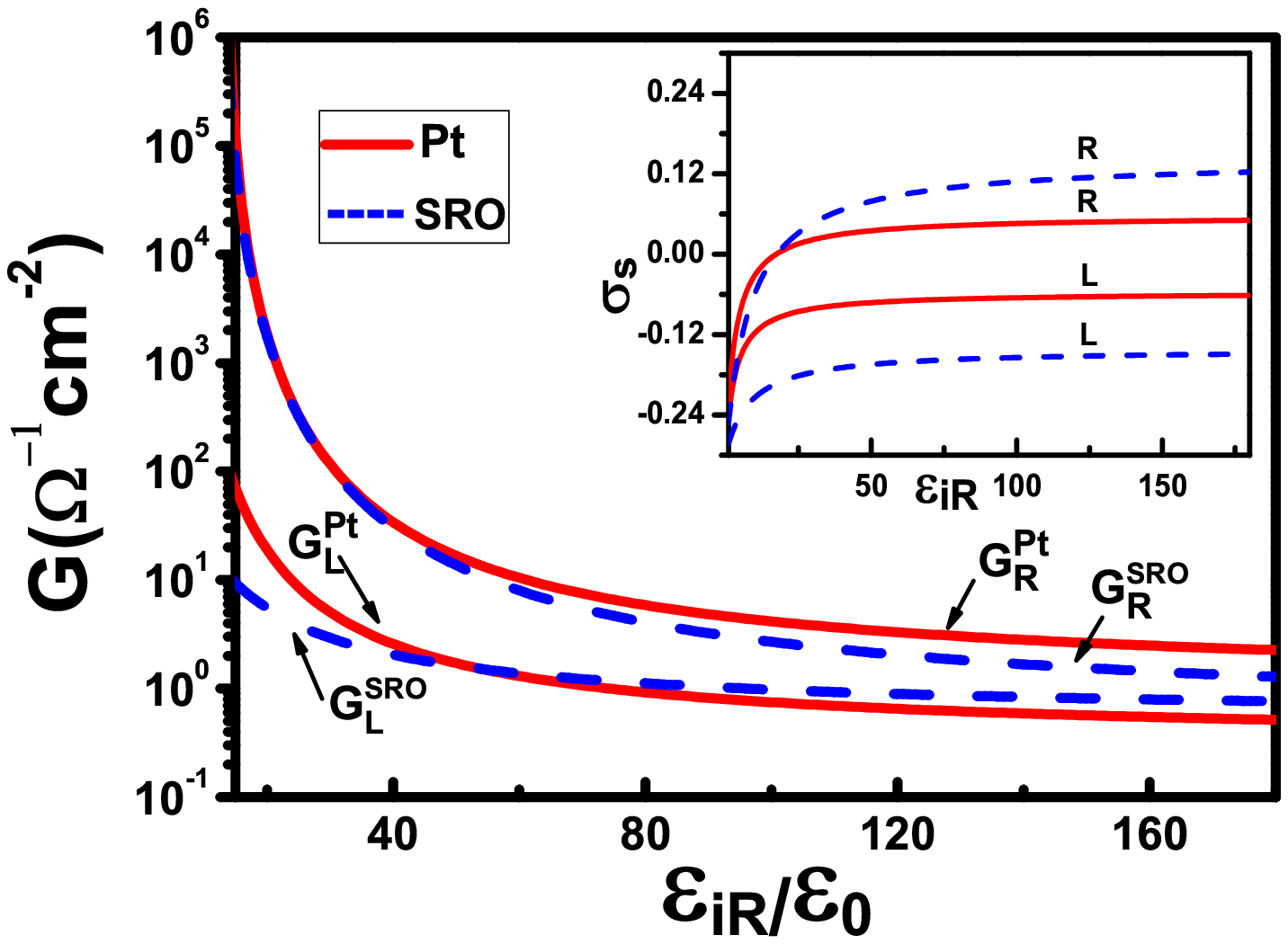}}
\caption{ \it{(Color online) TER as a function of (a)the dielectric constant of the left interface $\epsilon_{iL}$ and (b) the dielectric
constant of the right interface $\epsilon_{iR}$, and tunneling conductance as a function of (c) $\epsilon_{iL}$ and (d) $\epsilon_{iR}$. Solid
lines denote the case for Pt electrodes and dash lines for SRO electrodes. The inset in (c) and (d) represent the change of $\sigma_{S}$ with
$\epsilon_{iL}$ and $\epsilon_{iR}$ respectively. The letter L(R) in the insets stand for $P_{f}$ pointing to left(right). Here, the interface
polarization are taken as $|P_{iL}|=0.13C/m^2$ and $P_{iR}=-0.3C/m^2$.}} \label{Fig.3}
\end{figure}

To investigate the dependence of TER on the interfacial dielectric constants, polarization of the left and right interfaces are fixed as
$|P_{iL}|=0.13C/m^2$ and $P_{iR}=-0.3C/m^2$ respectively. The minus sign of $P_{iR}$ means that the pinned dipole $P_{iR}$ is assumed to point
to left. The above interface polarization are carefully chosen based on the first-principles calculations for the SRO/BTO/SRO
junction\cite{PRBLiu}.

 It is shown, from Fig.~3(a) and (b), that TER decreases
with the increasing of interfacial dielectric constants whether the corresponding interface dipole is reversal or not. Furthermore, TER varies
remarkably within the range of low $\epsilon_{iL}$ and $\epsilon_{iR}$, however, within the range of large interfacial dielectric constant, TER
changes slowly. Compared Fig.~3(a) with 3(b), we can obtain TER increases about three orders as $\epsilon_{iR}$ decreases, while TER is only
raised one order as $\epsilon_{iL}$ decreases. We also calculate TER for small magnitude of P$_{iR}$, and TER will increase slowly with the
decrease of $\epsilon_{iR}$. Therefore, the remarkable increase of TER when $\epsilon_{iR}$ decreases in Fig.~3(b) is due to the large quotient
$\frac{P_{iR}d_{iR}}{\epsilon_{iR}}$in Eq.~5, i.e., the great change of $\sigma_s$. From Fig.~3(a), one can also find that, compared with SRO
electrodes, TER is more sensitive to the variation of $\epsilon_{iL}$ for Pt electrodes(solid line). On the contrary, in Fig.~3(b), TER is more
sensitive to the change of $\epsilon_{iR}$ for SRO electrodes(dashed line). The above phenomena are caused by the variation of the distribution
of screening charges $\sigma_{s}$, which is shown in the insets of Fig.~3(c) and (d). In the inset of Fig.~3(c), $\sigma_{s}$ decreases when
$\epsilon_{iL}$ decreases for Pt electrodes when the barrier polarization points to left, and the decrease of $\sigma_{s}$ results in the
decrease of $G_L^{Pt}$ and then the steep increase of TER for a FTJ with Pt electrodes within low $\epsilon_{iL}$ range. With the decreasing of
$\epsilon_{iR}$ in the inset of Fig.~3(d), $\sigma_{s}$ decreases more quickly for SRO electrodes when the barrier polarization points to right
in comparison with other cases, which brings a steep increase of $G_R^{SRO}$ and TER in Fig.~3(d) and in Fig.~3(b), respectively. As effect of
electrodes on TER is concerned, SRO electrode is preferable to Pt electrode within the range of large $\epsilon_{iL}$ in Fig.~3(a), and for low
$\epsilon_{iL}$, Pt electrode is preferable. Thing will be completely opposite for the choice of electrode within the large and low
 scopes of $\epsilon_{iR}$ in Fig.~3(b).

In summary, the ferroelectric tunneling junction with symmetric electrodes is studied by using of Thomas-Fermi model and quantum tunneling
theory, and it is found that TER is induced in the above FTJ with a pinned interface dipole. TER as a function of the interface polarization and
dielectric constant is given, we obtain that TER can be great improved by enhancing the polarization of the switchable interface dipole, while
TER is not sensitive to the change of polarization of the nonswitchable interface dipole. A large TER is found when the nonswitchable interface
dipole points to the ferroelectric barrier. A significant TER can be obtained for low interface dielectric constants. The electrode effect on
TER is also presented. In one word, the model of a FTJ with only one pinned interface dipole is a practical model, and effects of the switchable
interface and the nonswitchable interface on TER are different. We hope our studies can provide suggestion and guide on the preparation of high
performance FTJs.

\begin{acknowledgments}
This work was supported by the National Science Foundation of China(Grant Nos.11047007 and 11174043), the QinLan project of Jiangsu Provincial
Education Committee. The author thanks Prof.~D.~Y.~Xing and Prof.~S.~Ju for helpful discussions.
\end{acknowledgments}

\newpage

Figures' Caption\\

FIG.~1: Sketch of the FTJ with a pinned interfacial dipole P$_{iR}$, whose direction can be arbitrarily preassigned. P$_{iL}$ and P$_{f}$ stand
for the polarization of the left interface and the medial ferroelectric film, and they can be reversed by an external electric field.  \\

FIG.2: (Color online) TER as a function of the magnitude of (a) $P_{iL}$ and (b) $P_{iR}$. Solid lines denote the TER for Pt electrodes, dashed
lines for SRO electrodes, and dash-dot line is drawn to indicate where the magnitude of the two interfaces are equal. Arrows
represent the direction of nonswitchable $P_{iR}$, which is assumed to preassign only one of the two directions.  \\

 FIG.3: (Color online) TER as a function of (a)the dielectric constant of the left interface $\epsilon_{iL}$ and (b) the dielectric constant
of the right interface $\epsilon_{iR}$, and tunneling conductance as a function of (c) $\epsilon_{iL}$ and (d) $\epsilon_{iR}$. Solid lines
denote the case for Pt electrodes and dash lines for SRO electrodes. The inset in (c) and (d) represent the change of $\sigma_{S}$ with
$\epsilon_{iL}$ and $\epsilon_{iR}$ respectively. The letter L(R) in the insets stand for $P_{f}$ pointing to left(right). Here, the interface
polarization are
taken as $|P_{iL}|=0.13C/m^2$ and $P_{iR}=-0.3C/m^2$.  \\

\end{document}